\documentclass[aps,prd,twocolumn,superscriptaddress,nofootinbib,showkeys]{revtex4}

\usepackage{graphicx}

\usepackage{euscript}

\usepackage{epsfig}

\usepackage{color}

\usepackage{amsfonts}

\usepackage{amsmath}

\usepackage{amssymb}

\begin{document}

\title{Appeasing the Phantom Menace?}

\author{Mariam Bouhmadi-L\'{o}pez}
\email{mariam.bouhmadi@ist.utl.pt} \affiliation{Centro
Multidisciplinar de Astrof\'{\i}sica - CENTRA, Departamento de
F\'{\i}sica, Instituto Superior T\'ecnico, Av. Rovisco Pais 1,
1049-001 Lisboa, Portugal}

\author{Yaser Tavakoli}
\email{tavakoli@ubi.pt} \affiliation{Departamento de F\'{\i}sica,
Faculdade de Ci\^encias -- UBI, Rua Marqu\^{e}s d'Avila e Bolama,
6200 Covilh\~{a}, Portugal}

\author{Paulo Vargas Moniz}
\email{pmoniz@ubi.pt} \affiliation{Departamento de F\'{\i}sica,
Faculdade de Ci\^encias -- UBI, Rua Marqu\^{e}s d'Avila e Bolama,
6200 Covilh\~{a}, Portugal} \affiliation{Centro Multidisciplinar de
Astrof\'{\i}sica - CENTRA, Departamento de F\'{\i}sica, Instituto
Superior T\'ecnico, Av. Rovisco Pais 1, 1049-001 Lisboa, Portugal}

\date{\today}

\begin{abstract}
An induced gravity brane-world model is considered herein. A
Gauss-Bonnet term is provided for the  bulk, whereas  phantom matter
is present on the brane. It is shown  that a combination of
infra-red  and ultra-violet modifications to general relativity
replaces a \emph{big rip } singularity: A \emph{sudden } singularity
emerges instead. Using current observational data, we also determine
a range of  values for the cosmic time corresponding to the sudden
singularity occurrence.

\end{abstract}

\date{\today}


\maketitle

\section{Introduction}

From current  observational data
\cite{Perlmutter:1998np,Spergel:2003cb,Cole:2005sx,Tegmark:2003uf},
it is now widely accepted that the universe is undergoing a state of
acceleration. The simplest setup to describe this acceleration is by
means of a cosmological constant, with an equation of state
$p_\Lambda=-\rho_\Lambda$, where $p_\Lambda$ is the pressure and
$\rho_\Lambda$ is the energy density of such a cosmological
constant. There are, however, other candidates that astronomical
observations still allow. Namely, a dark energy component
\cite{Perlmutter:1998np,Spergel:2003cb}  whose equation of state
parameter, $w$, is very close to $-1$, where $w$ is the ratio
between the pressure of dark energy, $p$, and its energy density
$\rho$. The relevance of this point is that dark energy can lead to
quite different scenarios concerning the future of the universe.  To
be more precise: if $w>-1$, dark energy corresponds to a
quintessence fluid \cite{Sahni:2002kh}; if $w=-1$, it is a
cosmological constant and the universe would be asymptotically de
Sitter; if $w<-1$, dark energy corresponds to a ``phantom'' content
\cite{Caldwell:2003vq}.

In the context of dark energy cosmology, the study of  gravitational
theories that entail spacetime singularities at late-times has made
a considerable progress in the last years. Being more specific, the
following cases have been established within a
Friedmann-Lema\^{\i}tre-Robertson-Walker (FLRW)
framework\footnote{For an alternative classification of the dark
energy related singularities see
Ref.~\cite{FernandezJambrina:2004yy}.} \cite{Nojiri:2005sx}
:

\begin{itemize}
\item Big rip singularity -  a singularity
at a finite cosmic time where the scale factor, the Hubble rate and
its cosmic time derivative diverge. \cite{Caldwell:2003vq,O1,O2};

\item Sudden singularity - a singularity
at a finite scale factor and in a finite cosmic time where the
Hubble rate is finite but its cosmic time derivative diverges
\cite{Barrow:2004xh,Nojiri:2004ip};

\item Big freeze singularity -  a singularity at a
finite scale factor and in a finite cosmic time where the Hubble
rate and its cosmic time derivative diverge \cite{BigFreeze};

\item Type IV singularity\footnote{Within the nomenclature of Ref.~\cite{Nojiri:2005sx}} - a singularity at a finite
scale factor and in a finite cosmic time where the Hubble rate and
its cosmic derivative are finite but higher derivative of the Hubble
rate diverges. These singularities can appear in the  framework
of modified theories of gravity \cite{Nojiri:2005sx}.
\end{itemize}

Subsequently, it has become of interest  to determine under which
conditions any of the above singularities can be removed, or at
least appeased in some manner. A significant range of  approaches
and corresponding results have been contributed to the  literature
\cite{BouhmadiLopez:2004me,Elizalde:2004mq,Sergei6,Dabrowski:2006dd,Kamenshchik:2007zj,BouhmadiLopez:2009pu,Sami:2006wj,scalarfield}.
Our purpose in this paper is to participate in  that line of
investigation, namely addressing the following question: Since the
big rip singularity occurs at high energy \emph{and} in the future,
could we expect that a combination of IR and UV effects removes or
replaces it? To this aim,  we employ in this paper
 a Dvali-Gabadadze-Porrati brane-world (DGP)
model \cite{DGP,Deffayet:2000uy} (see also \cite{sepangi}), that
includes a phantom matter fluid (that emulates the dark energy
dynamics), where the five-dimensional bulk is characterized by a
Gauss-Bonnet (GB) term. Generally speaking, within a DGP brane
component, we have a setting where infra-red
 (IR) effects modify general
relativity at late-time, whereas with the GB ingredient,
ultra-violet effects are present for high-energy scales
\cite{Brown:2005ug,otherGBIG,BouhmadiLopez:2008nf}.

Before proceeding into a more technical discussion, it could be of
interest to further add the following about having a phantom matter
fluid for the brane.  On the one hand, dark energy component with
$w<-1$, i.e. a phantom energy component has not yet been excluded by
the recent results of WMAP5. For example, the WMAP5 data (in
combination with other data) for a standard FLRW universe with
spatially flat sections, filled with cold dark matter and a dark
energy component with a constant equation of state parameter, $w$,
predicts $- 0.097<1+w<0.142$ for CMB, SNIa and BAO data which gives
the most stringent limit, while WMAP data alone predicts for this
model $-0.78<1+w<0.68$. For more details, see \cite{morewmap}. On
the other hand, to investigate future singularities,   a perfect
fluid is satisfactory and therefore we have not given an explicit
action for the phantom matter in terms of a minimally coupled scalar
field (with the opposite kinetical term) or through more general
scalar field actions like a k-essence action (see Section II). Let us also add that
it is well known that the DGP brane has a branch with an unstable
mode solution (i.e., a `ghost') \cite{Gregory:2008bf}. It may
therefore be questionable why to initiate a study within a DGP
setting or even insert phantom matter\footnote{For a
\emph{theoretical} criticism on phantom energy models, see however,
\cite{Carroll:2003st}.}. Our point is that, in spite of these open
lines, the features characterizing the DGP as well as GB elements
can provide a framework to investigate the intertwining of late time
dynamics and high energy effects.  An interesting ground to test it
is with a phantom fluid, since this matter in a standard FLRW
setting induces a big rip singularity. Eventually, the unresolved
issues for the DGP brane will be eliminated and results such as the
one we bring here will increase in interest.

This paper is then organized as follows. In Sect. II, we describe
our DGP-GB model filled with matter and a phantom fluid and present
the analytical solutions. In Sect. III, we analyze how the big rip
is replaced by a sudden singularity. Finally, we summarize our work
in Sect. IV, presenting also some possible lines to subsequently
investigate.

\vspace{0.5cm}

\section{The DGP-GB model with phantom matter}

The generalized Friedmann equation of a spatially flat DGP brane with a GB term in a Minkowski bulk
can be written as \cite{Brown:2005ug} (see also \cite{otherGBIG})
\begin{equation}
H^2=\frac{\kappa_5^2}{6r_c}\rho\pm
\frac{1}{r_c}\left(1+\frac83\alpha H^2\right)H,
\label{modifiedfriedmann}
\end{equation}
where   $\alpha$ is  the GB parameter,  $\kappa_5^2$ is the
five-dimensional gravitational constant and $r_c$ is the crossover
scale.  Moreover,  $\rho$ stands for the total energy density of the
brane. Therefore, for a late-time evolving brane the energy density
is well described by
\begin{equation}
\rho=\rho_b+\rho_{cdm}+\rho_{d},
\end{equation}
where $\rho_b$, $\rho_{cdm}$ and $\rho_{d}$ corresponds to the
energy density of baryons, cold dark matter and dark energy,
respectively. As $\rho_b$ and $\rho_{cdm}$ are both proportional to
$a^{-3}$, we will define their sum as $\rho_m$; i.e.
$\rho_m=\rho_b+\rho_{cdm}$. On the other hand, we will consider dark
energy to correspond to phantom energy. Finally, the total energy
density on the brane can be written as\footnote{Quantities with the
subscript $0$ denote their values as observed today.}
\begin{equation}
\rho=\frac{\rho_{m0}}{a^3}+\frac{\rho_{d0}}{a^{3(w+1)}},
\label{mattercontent}
\end{equation}
where $w,\rho_{m0},\rho_{d0}$ are constants and $1+w<0$.

It should
be noted that from (\ref{modifiedfriedmann}), we can obtain the
known \emph{self-accelerating} DGP solution \cite{Deffayet:2000uy}
($+$ sign in Eq.~(\ref{modifiedfriedmann}) with $\alpha = 0$), while the
\emph{normal} branch is retrieved for the $-$ sign with $\alpha =
0$; cf. \cite{BouhmadiLopez:2008nf} for more details and notation.

Let us then address  equation (\ref{modifiedfriedmann})
analytically,  selecting the $+$ sign,
adopting the approach presented in  \cite{BouhmadiLopez:2008nf}. This
equation with the matter content (\ref{mattercontent}) can be
expressed as
\begin{eqnarray}\label{eq5}
E^2(z)&=&\Omega_m(1+z)^3+\Omega_{d}(1+z)^{3(1+w)} \nonumber\\ &\,& +
2\sqrt{\Omega_{r_c}}\left [1+\Omega_\alpha E^2(z)\right ]E(z),
\end{eqnarray}
where $E(z) \equiv H/H_0$,  $z$ is the redshift and
\begin{eqnarray}\label{eq6}
 &\,&\Omega_m  \equiv \frac{\kappa_4^2 \rho_{m_0}}{3H_0^2},\,\,\,\,
\Omega_{d} \equiv \frac{\kappa_4^2\rho_{d_0} }{3H_0^2},\,\,\,\,\nonumber \\
&\,&\Omega_{r_c} \equiv \frac{1}{4r_c^2H_0^2},\,\,\,\,
\Omega_{\alpha} \equiv \frac{8}{3}\alpha H_0^2.
\end{eqnarray}

Evaluating the Friedmann equation (\ref{eq5}) at $z=0$ gives a
constraint on the cosmological parameter of the model
\begin{equation}
1=\Omega_{m}+\Omega_{d}+2\sqrt{\Omega_{r_{c}}}(1+\Omega_{\alpha}).\label{constraint}
\end{equation}
For $\Omega_{\alpha}=0$, we recover the constraint in the DGP model
without UV corrections. Coming back to our model, if we assume the
dimensionless crossover factor $\Omega_{r_{c}}$ to be  the same as
in the self-acceleration DGP model, then the similarities with a
spatially open universe are made more significant from the GB
effect, since $\Omega_{\alpha}>0$.

In order to obtain the evolution of the Hubble rate as a function of
the total energy density of the brane, we introduce the following
dimensionless variables:
\begin{eqnarray}\label{eq8}
\bar H&\equiv&\frac83 \frac{\alpha}{r_c}H =
2\Omega_\alpha\sqrt{\Omega_{r_c}}E(z), \label{eq8a}\\
\bar\rho&\equiv&\frac{32}{27}
\frac{\kappa_5^2\alpha^2}{r_c^3}
\rho\nonumber \\
&=&4\Omega_{r_c}\Omega_\alpha^2\left[\Omega_{d}(1+z)^{3(1+w)}+\Omega_m(1+z)^3\right],\label{dimensionlessrho}\\
b&\equiv&\frac83\frac{\alpha}{r_c^2}=4\Omega_\alpha\Omega_{r_c}.
\label{defb} 
\end{eqnarray}
In terms of these variables, the modified Friedmann equation then
reads:
\begin{equation}
{\bar H}^3-{\bar H}^2+b\bar H+\bar\rho=0. \label{Friedmannnb}
\end{equation}
Notice that there is a change of sign with respect to equation (17)
in \cite{BouhmadiLopez:2008nf}. The number of real roots is
determined by the sign of the discriminant function $N$ defined
as\footnote{The modified Friedmann equation (\ref{Friedmannnb}) will
be solved analytically following the method introduced in
Ref.~\cite{BouhmadiLopez:2008nf} for the normal DGP-GB branch. For a
semi-analytical approach see the second reference in
Ref.~\cite{Brown:2005ug}.} \cite{Abramowitz},
\begin{equation}\label{eq10}
N=Q^3+R^2,
\end{equation}
where $Q$ and $R$ are,
\begin{equation}\label{eq11}
Q=\frac{1}{3}\left(b-\frac{1}{3}\right),
~~~~~~~R=-\frac{1}{6}b-\frac{1}{2}\bar{\rho}+\frac{1}{27}.
\end{equation}
For the analysis of the number of physical solutions of the modified Friedmann
equation (\ref{Friedmannnb}), it is helpful to rewrite $N$ as
\begin{equation}\label{eq12}
N=\frac{1}{4}(\bar{\rho}-\bar{\rho}_{1})(\bar{\rho}-\bar{\rho}_{2}),
\end{equation}
where
\begin{equation}
\bar{\rho}_{1}=-\frac{1}{3}\left\{b-\frac{2}{9}[1+\sqrt{(1-3b)^3}]\right\},\label{rhoone}
\end{equation}
\begin{equation}
\bar{\rho}_{2}=-\frac{1}{3}\left\{b-\frac{2}{9}[1-\sqrt{(1-3b)^3}]\right\}.\label{rhotwo}
\end{equation}
Hence,  if $N$ is positive then there is a unique real solution. If
$N$ is negative, there are three real solutions, and finally, if $N$
vanishes, all roots are real and at least two are equal.


\begin{table*}
    \begin{tabular} 
{ | l | p{2.5cm} | l | p{3cm} |  p{3cm} |  p{3cm} |}
    \hline
    ~~~~~~~~b & $~~~~\bar{\rho_{1}}$ and $\bar{\rho_{2}} $  & ~~~~~~~Solutions for $\bar{H}$ & ~~~~~~$\eta$ and $\vartheta$  &~~~~Description \\ \hline 
    $\frac14\leq b<\frac13$ & $\bar{\rho_{1}}\leq0$,
    $\bar{\rho_{2}}<0$ & $\bar{H}_{1}=-\frac{1}{3}[2\sqrt{1-3b}\cosh(\frac{\alpha}{3})-1]$  &$\cosh(\alpha)\equiv-\frac{R}{\sqrt{-Q^3}}$
    $\sinh(\alpha)\equiv\sqrt{\frac{N}{-Q^3}}$, $\alpha_0<\alpha$, where $\cosh(\frac{\alpha_{0}}{3})=\frac{1}{2\sqrt{1-3b}}$.  & $\bar{H}_{1}<0$; contracting brane. \\ \hline
    $b=\frac13$ & $\bar{\rho_{1}}=\bar{\rho_{2}}=- \frac{1}{27}$ & $\bar{H}_{1}=-\frac{1}{3}[(1+27\bar{\rho})^{-\frac{1}{3}}-1]$ & &  $\bar{H}_{1}<0$; contracting brane. \\ \hline
    $\frac13<b$ & $\bar{\rho_{1}}$, $\bar{\rho_{2}}$:  complex
conjugates &
$\bar{H}_{1}=-\frac{1}{3}[2\sqrt{3b-1}\sinh(\frac{\vartheta}{3})-1]$
&$\cosh(\vartheta)\equiv\sqrt{\frac{N}{Q^3}}$ $\sinh(\vartheta)\equiv-\frac{R}{\sqrt{Q^3}}$, $\vartheta_0<\vartheta$, where $\sinh(\frac{\vartheta_{0}}{3})=\frac{1}{2\sqrt{1-3b}}$. & $\bar{H}_{1}<0$; contracting brane. \\
    \hline
    \end{tabular}
    \caption{Solutions for the algebraic equation (\ref{eq10}) with different ranges for $b$;
    See also Eqs.~(\ref{highenergy}) - (\ref{lowenergytwo}).}
    \label{table 1}
\end{table*}


An approximated bound for the value of $b$ can be established noticing
 that  $b$  is proportional to
$\Omega_{r_{c}}$ through (\ref{defb}). Hence, from the
equivalent quantity for the $\Omega_{r_{c}}$ in the DGP scenario
for the  self-accelerating branch \cite{Lazkoz-Maartens} and the
constraint on the curvature of the universe\footnote{At this respect, notice that at $z=0$, the term $2\sqrt{\Omega_{r_c}}(1+\Omega_{\alpha})$ mimics a curvature term on the modified Friedmann equation.} (see \cite{Spergel:2003cb}, e.g.),  its value should be small. These physical solutions can be included on the set of mathematical solutions with  $0<b<\frac{1}{4}$. 
Therefore, for the remaining of this
letter,  we shall study  in detail this setup. For completeness  the other
cases  are summarized in Table \ref{table 1}.

For
$0<b<\frac{1}{4}$ the values of $\bar{\rho}_{1}$ and $\bar{\rho}_{2}$ in equations
(\ref{rhoone}) and (\ref{rhotwo}) are real. More precisely, in this case
$\bar{\rho}_{1}>0$ and $\bar{\rho}_{2}<0$. The number of solutions of the cubic
Friedmann equation (\ref{Friedmannnb}) will depend on the values of the energy
density with respect to $\bar{\rho}_{1}$. As the (standard) energy density redshifts
backward in time (i.e. it  grows), we can distinguish three regimes: (i) high
energy regime: $\bar{\rho}_{1}<\bar{\rho}$, (ii) limiting regime:
$\bar{\rho}=\bar{\rho}_{1}$, (iii) low energy regime: $\bar{\rho}<\bar{\rho}_{1}$:

\begin{itemize}
\item During the high energy regime, the energy density of the brane is
bounded from below by $\bar{\rho}_{1}$. There is a unique solution for the
cubic Friedmann equation (\ref{Friedmannnb}), because
 $N$ is a positive function for this case. So the solution reads,
\begin{equation}
\bar{H}_{1}=-\frac{1}{3}\left[2\sqrt{1-3b}\cosh\left(\frac{\eta}{3}\right)-1\right],\label{highenergy}
\end{equation}
where $\eta$ is defined by
\begin{equation}\label{eq16}
\cosh(\eta)\equiv \frac{-R}{\sqrt{-Q^3}}~,~~~~~~~~\sinh(\eta)\equiv
\sqrt{\frac{N}{-Q^3}},
\end{equation}
and $\eta>0$. When $\eta\rightarrow0$, the energy density of the
brane approaches $\bar{\rho}_{1}$. This solution has a negative Hubble rate and
therefore it is unphysical for late-time cosmology.

\item During the limiting regime, $\bar{\rho}=\bar{\rho}_{1}$, the
function $N$ vanishes, and there are two real solutions
\begin{equation}
\bar{H}_{1}=-\frac{1}{3}\left(2\sqrt{1-3b}-1\right),\label{limitingone}
\end{equation}
\begin{equation}
\bar{H}_{2}=\frac{1}{3}(\sqrt{1-3b}+1).\label{limitingtwo}
\end{equation}
The solution $\bar{H}_{1}$ is negative and so it is also  not relevant
physically.

\item For the low energy regime,
$\bar{\rho}<\bar{\rho}_{1}$. In this case the function $N$ is
negative, and there are three different solutions. One of these
solutions is negative and corresponds to a contracting brane,
while the other two positive solutions correspond to  expanding
branes. Let us be more concrete:

The solution that describes the
contracting brane is similar to the corresponding solution of the
high energy regime:
\begin{equation}
\bar{H}_{1}=-\frac{1}{3}\left[2\sqrt{1-3b}\cos\left(\frac{\theta}{3}\right)-1\right],\label{lowenergy}
\end{equation}
where
\begin{equation}
\cos(\theta)\equiv\frac{-R}{\sqrt{-Q^3}}~,~~~~~~~~\sin(\theta)\equiv\sqrt{1+\frac{R^2}{Q^3}},\label{theta}
\end{equation}
and $0<\theta<\theta_{0}$. The parameter $\theta=0$ is defined as in equation
(\ref{theta}) in which the value of $\bar{\rho}$ reaches $\bar{\rho}_{1}$, and
the parameter $\theta_{0}$ corresponds to $\bar{\rho}=0$. For this  solution $\bar{H}_1$ is
negative and hence not suitable for the late-time cosmology. In addition, the solution approaches the same Hubble rate at
$\theta=0$ as
the limiting solution
(\ref{limitingone}).

\begin{figure*}
\includegraphics[width=1.0\textwidth]{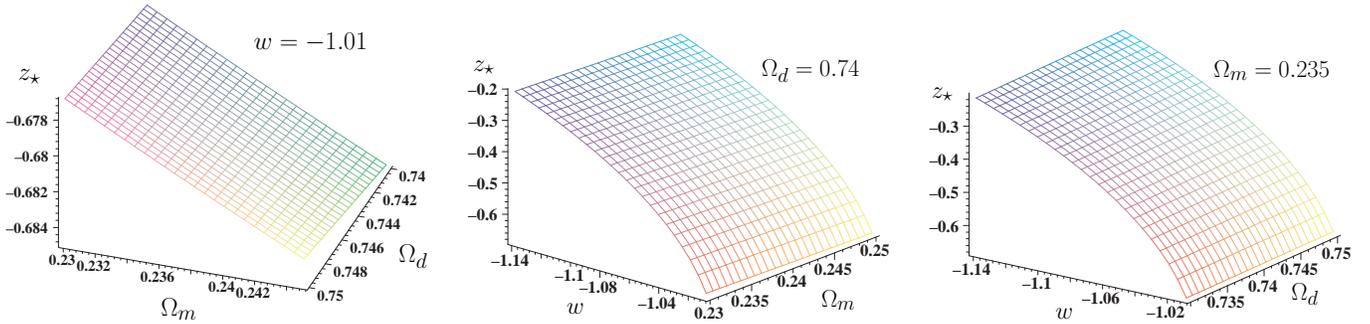}
\caption{Plot of the redshift  $z_{\star}$ at which the total energy density of the brane reaches its minimum value.}
\label{zstar}
\end{figure*}

On the other hand, the  two expanding branches are
described by
\begin{equation}
\bar{H}_{2}=\frac{1}{3}\left[2\sqrt{1-3b}\cos\left(\frac{\pi+\theta}{3}\right)+1\right],\label{lowenergyone}
\end{equation}
\begin{equation}
\bar{H}_{3}=\frac{1}{3}\left[2\sqrt{1-3b}\cos\left(\frac{\pi-\theta}{3}\right)+1\right],\label{lowenergytwo}
\end{equation}
For $\theta\rightarrow0$ the energy density $\bar{\rho}$ approaches
$\bar{\rho}_{1}$ and  the low energy regime connects at the limiting
regime with the solution (\ref{limitingtwo}) where both solutions coincide;
It can be further shown that $\bar{H}_{2}\leq \bar{H}_{3}$.
The more interesting solution for us is the brane expanding solution described by
Eq.~(\ref{lowenergyone}): This solution constitutes a
generalization of the  self-accelerating DGP solution with GB effects \cite{Brown:2005ug}, but
within  our herein\footnote{It should be clear that we have herein on
the paper a DGP-GB setting with phantom matter and therefore our
brane does not has, technically speaking, a self-accelerating phase,
asymptotically approaching a de Sitter stage in the future.} model.

\end{itemize}

\section{The brane escaping the big rip and plunging into a sudden singularity}

The usual self-accelerating DGP-GB solution is known to have a sudden
singularity in the \emph{past}, when the  brane is filled by standard matter
\cite{Brown:2005ug}. Herein, we consider instead  the brane filled with CDM
plus phantom energy, performing an analysis aiming at future cosmic time.
Starting from the modified Friedmann equation (\ref{Friedmannnb}), it  can be
shown that the first derivative of the Hubble rate is
\begin{equation}
\dot{H}=\frac{\kappa_{4}^2\dot{\rho}}{3[2H-\frac{1}{r_{c}}(1+8\alpha
H^2)]},\label{hderivative}
\end{equation}
where $\dot{\rho}$ is given by the equation of the energy conservation
\begin{equation}\label{eq23}
\dot{\rho}+3H[\rho_{m}+(1+w)\rho_{d}]=0,
\end{equation}
for the total energy density of the brane. Therefore, $\rho$ decreases initially until the redshift  reaches the value $z_{\star}$,
\begin{equation}
z_{\star}=-1+\left[-(1+w)\frac{\Omega_{d}}{\Omega_{m}}\right]^{-\frac{1}{3w}},\label{redshiftmin}
\end{equation}
where $\dot\rho=0$. Afterwards, the phantom matter starts dominating the expansion of the brane; indeed the brane starts super-accelerating\footnote{Notice that the denominator of equation (\ref{hderivative}) is always positive for the self-accelerating DGP-GB solution (\ref{lowenergyone}). This can be proven by using Eqs.~(\ref{hderivative}), (\ref{lowenergyone}) and the range of the variable $\theta$. Indeed, it can be shown that $0<3[2H-\frac{1}{r_{c}}(1+8\alpha
H^2)]<\frac{1-3b}{br_c}$. This means that in our model $\dot H$ is negative for $z_{\star}<z$ and positive for $z<z_{\star}$.} ($0<\dot{H}$) and the total energy density of the brane starts growing as the brane expands. In figure \ref{zstar}, we plot the redshift $z_{\star}$  versus the observational  values of the parameters\footnote{We estimate $z_\star$ in figure \ref{zstar} by  choosing  values for $\Omega_m$, $\Omega_d$ and $w$ in accordance with  the WMAP data \cite{Spergel:2003cb}. We know that those parameters are model dependent and therefore a  best fit analysis of the brane expansion using the currently available observational data (for example SNIa, BAO and CMB) would provide a much better analysis of $z_{\star}$. However, this analysis is far beyond the scope of the current work.} $\Omega_d$, $\Omega_m$ and $w$. As can be noticed in this figure, the total energy density of the brane would start increasing only in the future. Furthermore,  the larger  is $|w|$,  the sooner the phantom matter would dominate the expansion of the brane.  $\Omega_m$ has the opposite effect  while $\Omega_d$ has a much milder effect on  $z_{\star}$.

Substituting  Eq.~(\ref{eq23})
into Eq.~(\ref{hderivative}), the first derivative of the Hubble rate
reads,
\begin{equation}\label{eq24}
\dot{H}=-\frac{\kappa_{4}^2
H[(1+w)\rho_{d}+\rho_{m}]}{2H-\frac{1}{r_{c}}(1+8\alpha H^2)}.
\end{equation}
This equation shows that  when the Hubble rate approaches the constant
value\footnote{The Hubble rate given in Eq.~(\ref{eq25}) can be mapped into the dimensionless Hubble parameter  given in
Eq.~(\ref{limitingtwo}) for the limiting regime, reached at
the constant dimensionless energy density $\bar{\rho}=\bar{\rho}_{1}$.},
\begin{equation}\label{eq25}
H=\frac{r_{c}}{8\alpha}(\sqrt{1-3b}+1),
\end{equation}
the first derivative of the Hubble parameter, $\dot{H}$,
diverges, while
 the energy density of the brane remains finite.
Thus, instead of a scenario where  the energy density on the brane
blueshifts and eventually diverges, with a big rip singularity
emerging, we find that $(i)$  a finite value of the (dimensionless) energy density
$\bar{\rho}=\bar{\rho}_{1}$ (the same applies for the pressure), and $(ii)$ a finite value for the (dimensionless) Hubble
parameter $\bar{H}=\bar{H}_{2}$. Those values  are reached in the limiting regime described by Eq.~(\ref{limitingtwo}). And (iii) the first derivative of the Hubble parameter diverges at that point.
Therefore, the energy density is bounded, i.e. the limit
$z\rightarrow -1$ or $a\rightarrow \infty$ cannot be reached, and
instead  of a big rip singularity  we get  a sudden singularity,
despite the brane being filled with phantom matter.

\begin{figure}[h]
\includegraphics[width=0.8\columnwidth]{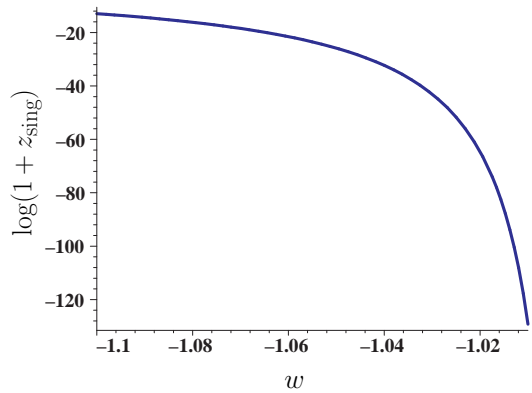}
\caption{Values for the redshift  at the sudden singularity,
$z_{\textrm{sing}}$, for fixed value of $\Omega_{d}=0.75$, $\Omega_m=0.24$ and $\Omega_{r_{c}}=5.95\times 10^{-6}$ with respect to the parameter $w$.}
\label{zsing}
\end{figure}

\begin{figure}[h]
\includegraphics[width=0.8\columnwidth]{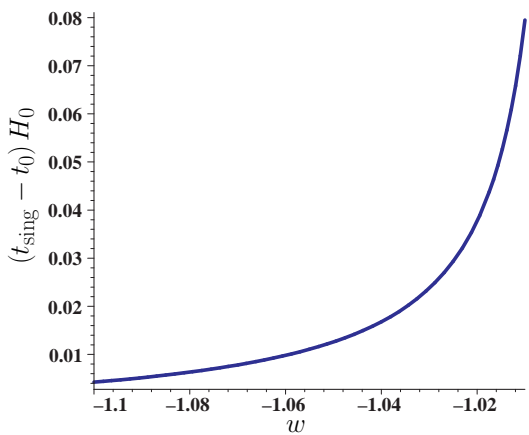}
\caption{Variation of the cosmic time left before the brane hits the sudden singularity,
$(t_{\rm{sing}}-t_0)H_0$ (in dimensionless units), for fixed values of $\Omega_{d}=0.75$, $\Omega_{m}=0.24$
and $\Omega_{r_{c}}=5.95\times 10^{-6}$ with respect to the equation of state
parameter $w$.}
\label{tsing}
\end{figure}

The results above can be further elaborated, by establishing
\emph{when} the sudden singularity  will happen (i.e., at which
redshift and cosmic time values):

\begin{itemize}

\item In order to obtain the redshift $z_{\textrm{sing}}$  where the sudden singularity takes place, we equate
  the dimensionless  total energy density of the brane (\ref{dimensionlessrho}) to its values at the sudden singularity  (cf. Eq.~(\ref{rhoone})). The allowed values for $z_{\textrm{sing}}$ are  plotted in Fig.~\ref{zsing} for the
various values of the equation of state parameter $w$.

\item Finally,
using the relation between the scale factor and the redshift
parameter, $a(t)=1/(1+z)$, one can write the Hubble rate as a
function of the redshift parameter and its cosmic time derivative as
follows

\begin{equation}\label{eq27}
H=\frac{\dot{a}}{a}=-\frac{\dot{z}}{1+z},
\end{equation}

and then integrating  this equation,  the cosmic time remaining before the brane hits the  sudden singularity
reads

\begin{equation} \label{t-end}
(t_{\textrm{sing}} - t_{0})H_0 = -
\int_{z=0}^{z=z_{\textrm{sing}}} \frac{dz}{(1+z)E(z)}.
\end{equation}

In the previous equation $t_{0}$ and $t_{\textrm{sing}}$ indicate the present time and
the time at the sudden singularity, respectively. We can plot
$t_{\textrm{sing}}$ for fixed values of $(\Omega_{m},
\Omega_{d},\Omega_{\alpha})$; See Fig.~\ref{tsing}. As we can notice
from Fig.~\ref{tsing}, the closer is the equation of state of the
phantom matter to that of a cosmological constant,  the farther
would be the sudden singularity. The same  plot is quite
enlightening as we can compare the age of the universe, essentially
$H_0^{-1}$, to the time left for such a sudden singularity to take
place on the future of the brane. As we can see, such a sudden
singularity would take place roughly in about 0.1 Gyr.

\end{itemize}

It is of interest to compare this estimate with the time at which a
big rip would take place in a standard four-dimensional universe filled with phantom matter whose equation of state is constant. In ref. \cite{O2}, the time remaining for such a
big rip singularity  was estimated to be about 22 Gyr. However, notice that in
\cite{O2}, $w=-1.5$ and $\Omega_m=0.3$ was used, whereas we have $w
\in [ -1.1, -1.01]$, $\Omega_m=0.24$;  Our setting is also
different. Within the framework of \cite{O2}, for $w=-1.1$ and
$\Omega_m=0.24$, one retrieves a time of about 80 Gyr. In
comparison, for the model presented in this paper, with $w=-1.5$ and
$\Omega_m=0.3$, $\Omega_d=0.68$, $\Omega_{\alpha}=0.005$, the
singularity emerges in $0.2$ Gyr. On the whole, although the
universe would meet a ``less severe'' singularity in the future, this
would happen much sooner that for the big rip case. In \cite{O2},
there are a few predictions concerning possible astronomical events
that would indicate the emergence of the singularity herein exposed.
Within our model, some of those events or other of similar impact
could occur rather earlier.

\section{Conclusions and Outlook}

The study of singularities occurring in the future is a subject of
interest and several proposals for their removal or substitution
have been advanced
\cite{BouhmadiLopez:2004me,Elizalde:2004mq,Sergei6,Dabrowski:2006dd,
Kamenshchik:2007zj,BouhmadiLopez:2009pu,Sami:2006wj,scalarfield}. In
this paper,  we have considered a specific case and investigated
whether a composition of specific IR and UV effects could alter a
big rip singularity setting. More concretely, we employed a simple
model: a DGP brane model, with phantom matter and a GB term for the
bulk. The DGP brane configuration has relevant IR effects, whereas
the GB component is important for high energies; phantom matter in a
standard FLRW model is known to induce the emergence of big rip
singularities \cite{Caldwell:2003vq}.

Our analysis indicates that the big rip can be  replaced  by a
sudden future singularity, through some intertwining between
late-time dynamics and high energy effects. Subsequently, we
determine values of the redshift and cosmic time,  before the brane
reaches the sudden  singularity. These results can be contrasted
with those, e.g., in \cite{O2} for the big rip
occurrence in a FLRW setting.

We are aware that the herein conclusions are based in a rather
particular result,  that was extracted from a specific model.
Subsequent research work would assist in clarifying some remaining
issues. For example, it would  be interesting to further study
if and how other singularities can be appeased or removed by means
of the herein combined IR and UV  effects. On the other hand, it might be 
interesting to consider  a modified Hilbert-Einstein action on the brane,
which in addition could alleviate the ghost problem present on the
self-accelerating DGP model by self-accelerating the normal DGP
branch \cite{BouhmadiLopez:2009db}, and see if some of the dark energy singularities can be removed or at least appeased in this setup. We hope to report on these lines in a forthcoming publication.


\section*{Acknowledgments}
MBL is supported by the Portuguese Agency Funda\c{c}\~{a}o para a
Ci\^{e}ncia e Tecnologia through the fellowships
SFRH/BPD/26542/2006. She also wishes to acknowledge the warm
hospitality of the Institute of  physics of the University of S\~ao
Paulo. YT is supported by the Portuguese Agency Funda\c{c}\~{a}o
para a Ci\^{e}ncia e Tecnologia through the fellowship
SFRH/BD/43709/2008. This research was supported by the grant
FEDER/POCI/FIS/P/57547/2004. The authors are grateful to S. Odintsov
for useful feedback.

\end{document}